# Explicit and implicit measures of emotions: Data-science might help to account for data complexity and heterogeneity

M.Moranges[a,b], C.Rouby[a], M.Plantevit[b,1], M.Bensafi[a,1,*]

[a] Lyon Neuroscience Research Center, CNRS, INSERM, Université Lyon1, France

[b] Laboratoire d'Informatique en Image et Systèmes d'information, Université Lyon1, CNRS, LIRIS UMR5205, F-69622, Lyon, France

## Abstract

Measuring emotions is a real challenge for fundamental and applied research, especially in ecological contexts. de Wijk and Noldus propose combining two types of measures - explicit to characterize a specific food, and implicit -physiological- to capture the whole experience of a meal in real-life situations. This raises several challenges including development of new and miniaturized sensors and devices but also developing new ways of data analysis. We suggest a path to follow for future studies regarding data analysis: to include Data Science in the game. This field of research may enable developing predictive but also explicative models that link subjective experience of emotions and physiological responses in real-life contexts. We suggest that food scientists should go out of their comfort zone by collaborating with computer scientists and then be trained with the new tools of Data Science, which will undoubtedly enable them 1/ to better manage complex and heterogeneous data sets, 2/ to extract knowledge that will be essential to this field of research.

## 1. Explicit and implicit measures of emotions

Capturing emotions with an objective point of view is a challenge for research, be it fundamental or applied to food. Two main trends have emerged: quantifying subjective events such as perceptions and hedonic judgments, that require introspection and eventually training, and measuring behavioral and/or physiological cues that are less accessible to awareness. The paper of de Wijk and Noldus (this issue) recommends a combination of both kinds of measurements: mainly explicit to characterize a specific food, mainly implicit to capture the whole experience of a meal or involuntary responses to a complex food. The authors underline the specificities and difficulties that pertain to each type of measurement.

Explicit measures provide precision on many subjective facets of perception, like valence and arousal, stimulation or relaxation and allow comparisons of foods along judgments of preference, familiarity, novelty, and precise profiling of a given food. The cognitive cost of such measurements is however high, because they need introspection and reflective thinking, which capture the end of the whole perception and evaluation processes. They give very limited access about dynamics, and no access to non-conscious events that are fast and fade quickly, but may weigh a lot in eating decisions.

On the other hand, implicit measurements follow processes that are automatic, too fast to reach awareness, that nevertheless influence decisions. Moreover, these recordings are dynamic, sensitive to both fast and basic physiological states and to slow conscious processes. They open a window on temporal processing of food perception, including the evolution of wanting and liking over time (Small et al., 2001). It is worth noting, however, that these measurements do not get direct access to the affective meaning of physiological events: they are clues, not a veridical reflection of a specific emotional processing. These clues are subject to variation between and within individuals, as attested by some discrepancies between studies. For example, whereas for some authors, the electrodermal response increases with the intensity of the stimulus (Bensafi et al., 2002b), for others it varies according to the unpleasantness (Brauchli et al., 1995) or olfactory quality (Møller & Dijksterhuis, 2003). Moreover, producing cognitive judgments like familiarity impairs some physiological responses: passive smelling and affective judgments of odors may evoke similar heart rate activity patterns, whereas familiarity judgment overwrites this autonomic response (Bensafi et al., 2002a). These differences between studies reflect a recurring phenomenon in physiological emotion studies: the great intra- and inter-individual diversity that characterizes the data sets generated by these studies. We will come back to this notion later.

## 2. Combining explicit and implicit measures of emotions in real-life situations

At the end of their paper, de Wijk and Noldus raise a challenge in the field of emotional science research: to use implicit measures in real life situations as well as real consumption situations in places like restaurants, stores, cinemas and others. The authors propose different methodological and scientific locks that will need to be lifted before such ecological measures can be initiated to monitor consumer experiences 24 h a day, 7 days a week. The first lock is the miniaturization of sensors and devices to measure psychophysiological responses such as electroencephalographic (EEG) responses, cardiac and respiratory responses, electrodermal response, etc. The second lock is the development of a new type of sensors to measure psychophysiological responses. Here, the new mobile devices and portable sensors have become powerful and non-invasive enough to allow such measurements to be considered in real conditions. The third lock that the authors highlight is new software that should facilitate the analysis of subjective, behavioral and physiological data in real-world conditions. It is on this third lock that we would like to provide a path to follow for future studies.

Indeed, if we embark on such measurements of food emotions in real-life situations, we will be able to perform long-term and very large-scale experiments but - at the same time - these experiments will produce a large amount of data. This will require the use of adequate techniques to be analyzed. de Wijk and Noldus highlight the complexity of the real-life experience and there is a real need to address the subject of processing these data, which will most certainly be numerous, complex, heterogeneous and highly subject to individual variability as mentioned earlier. Actually, implicit measures are sensitive to the effects of many factors (cognitive processing as stated above but also hunger, concordance between food and packaging, etc.). In order to arrive at an in-depth model of food emotions and their physiological basis in real situations one must i) evaluate the subjective and physiological responses in their heterogeneity and complexity by considering the multiple behavioral and psychophysiological parameters involved and ii) model in an intelligible way the rules linking these physiological parameters to the consumer's subjective emotional experience. Standard statistical approaches make it possible to address the problem of the relationship between the subjective experience of food-induced emotions and their physiological foundations on the basis of a limited number of parameters and comparisons (e.g. by comparing average responses for a known food vs. a novel food). However, EEG, psychophysiological, subjective and behavioral measures remain difficult to integrate into the same analysis. Furthermore, the wide inter- and intra-individual variability that characterizes studies of emotions (food and non-food) is difficult to consider as well, especially if sample sizes exceed the standards of conventional psychophysical studies. Here, we believe that integrating Data Science in the game might help on this front.

## 3. What is Data Science?

If you ask this question to different actors in the field, you are likely to receive different definitions. It is actually difficult to give a consensual definition of Data Science because many scientific fields claim it as an applied subfield of their discipline. However, one can define it on the basis of its motivations: leverage existing data sources and create new ones as needed in order to extract meaningful information and actionable insights. Such motivation is not novel and falls in the scope of research in Computer Science (e.g., artificial intelligence, machine learning, data mining, database theory) or Mathematics (e.g., Statistics). Its "novelty" (or success) relies on the fact that such motivation has become ubiquitous in many scientific domains and industrial applications fostering interdisciplinary research to produce new knowledge within a domain (Hey et al., 2009, Raghu and Schmidt, 2020), as for example food science.

Given that, such an approach borrows techniques rooted in Machine Learning, Data Mining and Statistics. Notice that the data preprocessing step should not be neglected (García et al., 2015), because it is as important as the analysis step since – if not correctly done – it can introduce important bias into the analysis and the interpretation of the results. This last point is particularly important in studies on food emotions that combine explicit measures (e.g. from subjective ratings to questionnaire data) and implicit measures (e.g. from single physiological parameters to large time-series) because the volume and the format of the data are very heterogeneous from one type of data to another, especially if the context of data acquisition is not well controlled (e.g. real-life studies).

## 4. How Data Science can be applied to food emotion studies?

Among the various data science analysis tasks that can be used in food emotion studies in real-life context, two important and complementary tasks can be distinguished:

1/ Global predictive modeling (Alpaydin, 2020), which aims to take advantage of multifaceted data to learn and construct an emotion prediction model based on different physiological measures. The ultimate goal of such a task is the automation of a process, e.g. automatically detecting whether a food evokes pleasure based on multimodal physiological responses (EEG responses, heart rate, electro-dermal response etc.) but also on factors intrinsic to the participant (e.g. satiety, knowledge about the food etc.) or extrinsic variables (e.g. context, real-life environment etc.). Such modeling emphasizes the predictive power of the model to the detriment of its interpretability (explanatory power). In other words, we will be able to predict emotion on the basis of physiology but without necessarily explaining the mechanisms underlying this relationship.

2/ Exploratory analysis does not aim to predict but, instead, has a strong explanatory power. This form of descriptive modeling aims to automatically discover new information about the domain (e.g. food emotions) in which the data were measured. To this end, machine discoveries are used to synergistically boost expertise of the food scientist. Here, particular attention is paid to interpretability of the results. In this domain, Subgroup Discovery (SD) (Atzmueller, 2015) (aka Exceptional Model Mining, EMM (Duivesteijn et al., 2016) makes it possible to discover subgroups (subsets) within the data that significantly deviate from the whole data (e.g., finding subgroups of individuals who have a common and particular physiological response to pleasant foods). These subgroups are identified by conditions on some feature combinations of physiological measures and are easily understandable.

One can note that making accurate predictions remains difficult in the context of human cognition, especially because this requires one to consider the intrinsic inter-individual variability of human emotions. On this point, Exploratory analysis can help to elicit new hypotheses by analyzing together both implicit and explicit measures. Especially, Subgroup discovery makes it possible to automatically discover rules between multiple implicit (physiological) measures and explicit measures for a specific subpopulation (e.g., gender = female, age <40). This type of analysis, which makes it possible to characterize subsets of individuals, can be of definite help not only to food scientists to help them better understand the diversity underlying the psychology and physiology of food emotions, but also to industrialists in the field so that their new products are more in line with the emotional expectations of specific populations of consumers. In fact, the boundary between Global predictive modeling and Exploratory analysis is not as clear-cut as we present it. New horizons are opening up in this field, in particular the field of explainable artificial intelligence (xAI) for which one expects explanation about the decision made by a black box model (e.g., deep neural networks) (Guidotti et al., 2018, Marcinkevičs and Vogt, 2020) to also support reproducibility and replicability (National Academies of Sciences et al., 2019). Thus, developing models that predict food emotion on the basis of multiple physiological measurements obtained in real-life conditions, and especially explaining the mechanisms underlying this relationship is a realistic challenge.

## 5. Conclusions

In conclusion, we are in total agreement with de Wijk and Noldus to consider that combining explicit and implicit measures is a necessity if we want to better understand the food emotions evoked in real life situations. Explicit and implicit measures are complementary and will likely enable approaching access to behavioral and physiological underpinnings of emotions. The problem is whether we really benefit from this combination of data with classical statistical approaches. Here, we suggest that the food scientist must go out of his/her comfort zone by collaborating with computer scientists and then be trained in the new tools of Data Science, which will undoubtedly enable him/her 1/ to better manage complex and heterogeneous data sets, 2/ to extract knowledge that will be essential to this field of research.

# Funding

This study was granted by the ANR ChemoSim project.

# CRediT authorship contribution statement

**M. Moranges:** Conceptualization, Methodology, Writing - original draft, Writing - review & editing. **C. Rouby:** Conceptualization, Methodology, Writing - original draft, Writing - review & editing. **M. Plantevit:** Conceptualization, Methodology, Writing - original draft, Writing - review & editing. **M. Bensafi:** Conceptualization, Methodology, Writing - original draft, Writing - review & editing, Funding acquisition, Project administration.

# Declaration of Competing Interest

The authors declare that they have no known competing financial interests or personal relationships that could have appeared to influence the work reported in this paper.

# Acknowledgments

We wish to thank the ANR for their support (ANR-ChemoSim project).